\documentclass[11pt]{article}

\usepackage{amssymb}
\usepackage{amsmath}
\usepackage{graphicx}

\def\beq{\begin{eqnarray}}    
\def\eeq{\end{eqnarray}}      

\def\al{\alpha}
\def\be{\beta}

\def\ze{\zeta}

\def\ka{\kappa}
\def\La{\Lambda}

\def\La{\Lambda}

\def\Om{\Omega}

\newcommand{\OM}{\Omega_M}
\newcommand{\OL}{\Omega_{\Lambda}}

\newcommand{\CC}{\Lambda}

\begin{document}


\newpage


\begin{center}
{\large \textsc{Variable Cosmological Constant}} \vskip 2mm

{\large \textsc{as a Planck Scale Effect}} \vskip 8mm


\textbf{Ilya L. Shapiro,} \footnote{ On leave from Tomsk State
Pedagogical University, Tomsk, Russia.\quad E-mails address:
shapiro@fisica.ufjf.br} $\,\,$ \textbf{Joan Sol\`{a}}
\footnote{On leave of absence from the Dept. E.C.M. and C.E.R  for
Astrophysics, Particle  Physics and Cosmology,\\
\mbox{\hspace{0.5cm}} Universitat de
 Barcelona, Diagonal 647, E-08028,
Barcelona, Spain. E-mail address: sola@ifae.es} \vskip 3mm

\textsl{Departamento de F\'{\i}sica, ICE, Universidade Federal de
Juiz de Fora, MG, Brazil}

\textsl{Juiz de Fora, 36036-330, Minas Gerais, Brazil}

\vskip10mm

\textbf{Cristina Espa\~{n}a-Bonet, \   Pilar Ruiz-Lapuente}

\vskip2mm

 \textsl{Departament of Astronomy, University of Barcelona,
\thinspace \\
\textsl{ Diagonal 647, E-08028, Barcelona, Spain}}

\textsl{CER for Astrophysics, Particle Physics and Cosmology, U.
Barcelona, Spain\\}

\end{center}
\vskip 15mm

\begin{quotation}
\noindent {\large\it Abstract}.$\,\,$ We construct a semiclassical
Friedmann-Lema\^\i tre-Robertson-Walker (FLRW) cosmological model
assuming a running cosmological constant (CC). It turns out that
the CC becomes variable at arbitrarily low energies due to the
remnant quantum effects of the heaviest particles, e.g. the
Planck scale physics. These effects are universal in the sense
that they lead to a low-energy structure common to a large class
of high-energy theories. Remarkably, the uncertainty concerning
the unknown high-energy dynamics is accumulated into a single
parameter $\,\nu$, such that the model has an essential
predictive power. Future Type Ia supernovae experiments (like
SNAP) can verify whether this framework is correct. For the flat
FLRW case and a moderate value $\nu\sim 10^{-2}$, we predict an
increase of $10-20\%$ in the value of $\Omega_{\Lambda}$ at
redshifts $z=1-1.5$ perfectly reachable by SNAP.
\end{quotation}
\vskip 8mm

\newpage

\vskip 6mm

 \noindent {\bf Introduction}\quad

 \vskip 0.4cm

The number and versatility of publications concerning the
Cosmological Constant (CC) problem, show that it is a fascinating
interface between Cosmology and Quantum Field Theory (QFT). One
can identify three distinct aspects of this problem: $\,$ I) The
famous ``old'' CC problem (see \cite{weinRMP}) is why the induced
and vacuum counterparts of the CC cancel each other with such a
huge precision; $\,$ II) The coincidence problem (see the reviews
\cite{reviews}) is why the observed CC in the present-day
Universe \cite{Supernovae} is so close to the matter density
$\,\rho\,$; $\,$ III) The dynamics responsible for the CC
problem: is the CC some form of dark energy associated to a
special new entity?  If future astronomical observations (like
the SNAP project \cite{SNAP}) would discover that the CC depends
on the redshift parameter $z$, one could wonder whether this
variation can be achieved only through the mysterious
quintessence scalar field searched by everyone. The idea of a
scalar field that may adjust itself such that the cosmological
constant is zero or very small, has been entertained many times
in the literature and since long ago \,\cite{Dolgov,PSW}. But in
practice one ends up with all sorts of (more or less obvious)
fine-tunings. As emphasized in \cite{weinDM,nova,nova2}, the
quintessence proposal for problem III does not help much in
solving problems I and II. In our opinion, before introducing an
\textit{ad hoc} field we had better check whether the variation
of the CC can be attributed to a simpler, more economical and
truly robust QFT concept, like e.g. the renormalization group
(RG).

The relation between the RG and the CC problem had been in air for
a long time \cite{polyakov-CC,lam}, but only recently it has been
settled into the phenomenological framework
\cite{nova,nova2,babic,Reuter03a}. A consistent formulation of the
approach within the context of particle physics in curved
space-time has been presented in \cite{nova2}, where the relevant
RG scale $\,\mu\,$ has been identified with the energy of the
gravitational quanta; that means, in the cosmological arena, with
the Hubble parameter: $\mu=H$. Other possibilities have been
considered in the literature and in different contexts
\,\footnote{In Ref.\cite{nova} the alternative possibility
associating $\mu$ with $\rho^{1/4}$ is put forward, and in
\cite{babic} it is amply exploited. In Ref.\cite{Reuter03a} the
RG scale $\mu$ was identified with the inverse of the age of the
universe at any given cosmological time, i.e. $\mu\sim 1/t$.  This
is essentially equivalent to our choice\,\cite{nova2}, because
$H\sim 1/t$ in the FLRW cosmological setting.}. At first sight,
the tiny present-day value $\,H_0\sim 1.5\times 10^{-42}\,GeV$
makes the RG running of the CC senseless, because all the massive
fields should decouple. Of course, this is true if we apply the
``sharp cut-off'' scheme of decoupling \cite{nova}. However, in
\cite{babic} a pertinent observation concerning the decoupling
has been made. In fact, the decoupling is not sharp, and the
quantum effects of a particle with mass $m$ at the energy $\mu
\ll m$, even if suppressed by the factor $\,(\mu / m)^2$, can be
important in the CC context. Recent investigations of the
decoupling in the gravitational theory \cite{apco} have confirmed
explicitly this decoupling behavior for the higher-derivative
sector of the vacuum action. However, the running of the CC and
the inverse Newton constant is not visible within the
perturbative approach of \cite{apco}, while alternative covariant
methods of calculations are incompatible with the mass-dependent
renormalization scheme and their development may require a long
time. In this situation it is reasonable to apply the
phenomenological approach. In the present paper we suppose that
the decoupling of the CC occurs in the same way as in those
sectors of the vacuum action, where we are able to obtain this
decoupling explicitly \cite{apco}. As we are going to show, this
approach may shed light to the coincidence problem II and leads
to very definite predictions for the variable CC, which can be
tested by the rich program of astronomical observations scheduled
for the next few years.

\vspace{0.7cm}

\noindent {\bf Renormalization Group and Planck Scale
Physics}\quad

\vspace{0.4cm}

Consider a free field of spin $J$, mass $M_J$ and multiplicities
$(n_{c}$, $\,n_{J})$ in an external gravitational field --  e.g.
$\,(n_{c},n_{1/2})=(3,2)\,$ for quarks, $(1,2)$ for leptons and
$\,(n_{c},n_{0,1})=(1,1)\,$ for scalar and vector fields. At high
energy scale, the corresponding contribution to the $\be$-function
$d\Lambda/d\ln\mu$ for the CC is \cite{nova2}
\begin{equation}
\beta_{\Lambda}(\mu \gg M_J)
\,\,=\,\,\frac{(-1)^{2J}}{(4\pi)^2}\,(J+1/2)
\,n_{c}\,n_{J}\,\,\,M_{J}^{4}\,. \label{beta lambda UV}
\end{equation}
At low energies ($\mu \ll M_J$) this contribution is suppressed
due to the decoupling. At an arbitrary scale $\,\mu\,$ (whether
greater or smaller than $\,M_{J}\,$), the contribution of a
particle with  mass $\,M_{J}\,$ should be multiplied by a form
factor $\,F(\mu / M_J)\,$. At high energies $\,F(\mu \gg
M_J)\simeq 1\,$ because there must be correspondence between the
minimal subtraction scheme and the physical mass-dependent schemes
of renormalization at high energies. In the low-energy regime
$\,\mu \ll M_J\,$ one can expand the function $\,F$ into powers
of $\mu/M_J$:
\begin{equation}
F\Big(\frac{\mu}{M_J}\Big) \,=\,\sum_{n=1}^\infty
k_n\Big(\frac{\mu}{M_J}\Big)^{2n}\,. \label{function}
\end{equation}
Two relevant observations are in order. First, the term $n=0$ must
be absent, because it would lead to the non-decoupling of $M_J$,
with untenable phenomenological implications on the CC value.
Indeed, for those terms in the vacuum action where the derivation
of the function $\,F(\mu / M_J)\,$ is possible \cite{apco}, the
$\,n=0\,$ terms are really absent. Second, due to the covariance
the number of metric derivatives (resulting into powers of $H$)
must be even, and so there are no terms with odd powers in the
expansion (\ref{function}) -- remember we identify $\mu$ with the
Hubble parameter. No other restrictions for the coefficients
$\,k_n\,$ can be seen. Indeed, in the $\,H \ll M_J\,$ regime the
most relevant coefficient is $\,k_1$. In the rest of this article
we develop a cosmological model based on the hypothesis that this
coefficient is different from zero. This phenomenological input
does not contradict any known principle or law of physics and at
the present state of knowledge its validity can be checked only by
comparison with the experimental data.

As far as we suppose the usual form of decoupling for the CC, all
the contributions in (\ref{beta lambda UV}) are suppressed by the
factor of $\,(\mu / M_J)^2\,$ and we arrive at the expression
\begin{equation}
\beta_{\Lambda}(\mu \ll M_J) \,\,\simeq k_1
\,\,\frac{(-1)^{2J}}{(4\pi)^2}
\,(J+1/2)\,n_{c}\,n_{J}\,\,M_{J}^2\,\mu^2\,. \label{beta lambda
IR}
\end{equation}
At the very low (from the particle physics point of view) energies
$\,\mu = H_0\sim 1.5 \times 10^{-42}\,GeV$, the relation $\,(\mu /
M_J)^2\ll 1$ is satisfied for all massive particles: starting from
the lightest neutrino, whose presumed mass is $\,m_\nu \approx
10^{30}\,H_0$, up to the unknown heaviest particle $M_+\lesssim
M_P$. Then, according to (\ref{beta lambda IR}), the total
$\be$-function for the CC in the present-day Universe is, in very
good approximation, dominated by the heaviest masses:
\begin{equation}
\beta_{\Lambda}^t
\,=\,\sum_{M_J=m_\nu}^{M_+}\,\beta_{\Lambda}(M_J) \,\simeq
\,\frac{1}{(4\pi)^2}\,\,\sigma\,M^2\,\mu^2\,. \label{beta lambda}
\end{equation}
Here we have introduced the following parameters: $\,M\,$ is a
mass parameter which represents the main feature of the total
$\be$-function for the CC at the present cosmic scale, and
$\sigma=\pm 1$ indicates the sign of the CC $\be$-function,
depending on whether the fermions ($\sigma=-1$) or bosons
($\sigma=+1$) dominate at the highest energies. The quadratic
dependence on the masses $M_{J}$ makes $\beta_{\Lambda}^t$ highly
sensitive to the particle spectrum near the Planck scale while the
spectrum at lower energies has no impact whatsoever on
$\beta_{\Lambda}^t$.

Having no experimental data about the highest energies, the
numerical choice of $\,\sigma\,M^2\,$ is model-dependent. For
example, the fermion and boson contributions in (\ref{beta
lambda}) might cancel due to supersymmetry (SUSY) and the total
$\be$-function becomes non-zero at lower energies due to SUSY
breaking. In this case, the value of $M^2$ depends on the scale of
this breaking, and the sign $\sigma$ depends on the way SUSY is
broken. In particular, the SUSY breaking near the Fermi scale
leads to a negligible $\beta_{\Lambda}^t$, while the SUSY breaking
at a GUT scale (particularly at a scale near the Planck mass)
provides a significant $\beta_{\Lambda}^t$.

Another option is to suppose some kind of string transition into
QFT at the Planck scale. Then the heaviest particles would have
the masses comparable to the Planck mass $M_P$ and represent the
remnants, e.g., of the massive modes of a superstring. Of course,
this does not contradict SUSY at the lower energies, when these
string WIMPzillas decouple from the matter sector. But, according
to Eq. (\ref{beta lambda}), they never decouple completely from
the RG in the CC case. Below, we shall work with this option and
take, for simplicity, $M^2=M_P^2$. Let us clarify that this choice
does not necessary mean that the relevant high energy particles
have the Planck mass. The mass of each particle may be indeed
smaller than $M_P$, and the equality, or even the effective value
$M\gtrsim M_P$, can be achieved due to the  multiplicities of
these particles. With these considerations in mind, our very first
observation is that the natural value of the $\beta$-function
(\ref{beta lambda}) at the present time is
\begin{equation}
\Big|\beta_{\Lambda}^t\Big|
\,=\,\frac{1}{(4\pi)^2}\,\,M^2\,\cdot\,\mu^2
\,=\frac{c}{(4\pi)^2}\,M_P^2\,\cdot\,H_0^2\sim 10^{-47}\,GeV^4\,,
\label{beta Planck}
\end{equation}
where $c$ is some coefficient. For $c={\cal O}(1-10)$ the
$\,\beta_{\Lambda}$ function is very much close to the
experimental data on the CC \cite{Supernovae}. This is highly
remarkable, because two vastly different and (in principle)
totally unrelated scales are involved to realize this
``coincidence'': $H_0$ (the value of $\mu$ at present) and $M_P$,
being these scales separated by more than $60$ orders of
magnitude! In this framework the energy scale of the CC is of the
order of the geometrical mean of these two widely different ones,
and only near $M_P$ all scales become of the same order. Moreover
at low energy the running of the CC proceeds smoothly and in the
right ballpark required by the natural solution of problem II. In
fact, this is not a complete solution of this problem. Of course
Eq. (\ref{beta lambda}) holds at any cosmic scale, but the
Friedmann equation leads to the relation $H \sim \sqrt{\La}/M_P$
only at lower energies, while at higher energies the matter or
radiation density dominates. Anyway, the coincidence between the
supposed RG decoupling (\ref{beta lambda}) and the Friedmann
equation for the modern, CC-dominated, Universe looks rather
intriguing and is worth exploring.

\vspace{0.7cm}

 \noindent {\bf A cosmological FLRW model with running $\Lambda$}\quad

 \vspace{0.4cm}

Consider the implications for FLRW cosmological models coupled to
Eq. (\ref{beta lambda}). The first step is to derive the CC
dependence from the redshift parameter $z$, defined as
$1+z=a_0/a$, where $a_0$ is the present-day scale factor. Using
the identification of the RG scale $\mu$ with $H$, we reach the
equation
\begin{equation}
\frac{d\La}{dz}\,=\,\frac{1}{H}\,\frac{dH}{dz}\,
\beta_{\Lambda}^t\,=\,\frac{\sigma\,M^2}{(4\pi)^2}\,H\,\frac{dH}{dz}\,.
\label{beta z}
\end{equation}
In order to construct the cosmological model, we shall use, along
with Eq. (\ref{beta z}), the Friedmann equation
\begin{equation}
H^{2}\equiv \left( \frac{\dot{a}}{a}\right) ^{2}=\frac{8\pi G }{3}
\left( \rho +\Lambda\right) -\frac{k}{a^{2}}\,, \label{Friedmann}
\end{equation}
where $\rho$ is the matter/CDM/radiation density and the last
curvature-dependent term can be presented as
$\,\,-k/a^2=H_0^2\,\Omega_K^0(1+z)^2$, where $\Omega_K^0$ is the
spatial curvature parameter at present ($z=0$), and can be written
in terms of the usual cosmological parameters for matter and CC at
the present time: $\Omega_K^0=1-\Omega_M^0-\Omega_{\Lambda}^0$.
Also, the energy conservation law provides the third necessary
equation \beq
\frac{d\La}{dt}\,+\,\frac{d\rho}{dt}\,+\,3H\,(\rho+p)\,=\,0\,,
\label{conservation} \eeq where $p$ is the matter/radiation
pressure. One could wonder whether the non-local effects behind
the renormalization group are compatible with the standard energy
conservation equation (\ref{conservation}). Let us remember,
however, that the covariant form of the conservation law
$\,<\nabla_\mu T^\mu_\nu > =0\,$ just reflects the covariance of
the effective action and therefore does not depend on the
non-localities which are always present in the quantum
corrections. Furthermore, in a situation where the energy scale
associated to the metric derivatives is very small, the leading
effect of the CC scale dependence may be, according to our model,
presented in a compact form due to the renormalization group.
Then the proper form of the conservation law is the one we use
here.

As we shall consider both MD (matter dominated) and RD (radiation
dominated) regimes, it is useful to solve the equations
(\ref{beta z}), (\ref{Friedmann}), (\ref{conservation}) using an
arbitrary equation of state $\,p=\al\rho$, with $\alpha=0$ for MD
and $\alpha=1/3$ for RD. The time derivative in
(\ref{conservation}) can be easily traded for a derivative in
$\,z\,$ via $d/dt=\,-\,H\,(1+z)\,d/dz$. Hence we arrive at a
coupled system of ordinary differential equations in the $z$
variable. The solution for the matter-radiation energy density and
CC is completely analytical. It reads a follows:
\begin{equation}
\rho(z;\nu)
\,=\,\Big(\rho_0+\frac{\ka}{\ze-2}\,\rho_c^0\Big)\,(1+z)^{\ze}
-\frac{\ka}{\ze-2}\,\rho_c^0\,(1+z)^2 \label{rho}
\end{equation}
and
\begin{equation}
\Lambda(z;\nu)=\Lambda_0+\frac{3\,\nu}{8\,\pi}\,M_P^2\left(H^2(z;\nu)-H_0^2\right)=
\Lambda_0+\rho_0\,f(z)\,+\rho_c^0\,g(z)\,, \label{Lambda 1}
\end{equation}
with
\begin{eqnarray}
f(z)=\frac{\nu}{1-\nu}\,\left[\left(1+z\right)^{\ze}-1\right]\,,
\label{f}
\end{eqnarray}
\begin{eqnarray}
g(z)\,\,=-\frac{\ka\,(1+3\al)}{2\,(\ze-2)}\,z\,(z+2)
+\,\nu\,\ka\,\frac{\left(1+z\right)^{\ze}-1}{(1-\nu)(\ze-2)}\,.
\label{g}
\end{eqnarray}
Here $\rho_0$, $\rho_c^0$, $\Lambda_0$ and $H_0$ are respectively
the matter-radiation energy density,  critical density, CC and
Hubble parameter at present, and we have used the following
dimensionless coefficients:
\begin{equation}
\nu\,=\,\frac{\sigma\,M^2}{12\pi M_P^2} \,,\,\,\,\,\,\,\,\,\,\,
\ze\,=\,3\,(1-\nu)\,(\al+1) \,,\,\,\,\,\,\,\,\,\,\,
\ka\,=-2\,\nu\,\Om_K^0\,. \label{notations 1}
\end{equation}
It should be clear that there is only one single independent
parameter in the model: $\nu$. In the limit $\nu\rightarrow 0$ we
recover the standard result for $\,\rho(z)\,$ with constant CC
(see, e.g. \cite{GC}). In order to avoid confusion, let us note
that the above solutions for $\,\rho(z;\nu)$ and $\,\La(z;\nu)\,$
have no singularity in the limits $\,\ze=2\,$ and $\,\nu = 1$. It
is easy to see that the function $\,g(z)\,$ differs from zero
only for the cases of non-vanishing spatial curvature $\,\ka\neq
0$, while another function $\,f(z)\,$ remains significant even
for the flat space case $\ka=0$. Furthermore, in the limit $\,\nu
\to 0\,$, $\Lambda(z;\nu)$ just becomes the $z$-independent
standard CC.

The formulas above represent the universal solution at low
energies, when all massive particles decouple according to
(\ref{beta lambda IR}). In order to understand the limits of
applicability of these solutions, let us notice that when the
temperature of the radiation energy density achieves the Fermi
scale $\,T \sim M_F \simeq 293\,GeV$, the value of the Hubble
parameter is just $\,H(M_F)\equiv 1.2\times 10^{-4}\,eV$, unable
to excite even the lightest neutrino. However, the most
interesting scales are much smaller. Let us first consider the
nucleosynthesis epoch when the radiation dominates over the
matter, and derive the restriction on the parameter $\,\nu$. In
the RD regime, the solution for the density (\ref{rho}) can be
rewritten in terms of the temperature
\begin{eqnarray}\label{rhozR2}
\rho_R(T)=\frac{\pi^2\,g_{\ast}}{30} \,\left(r^{-\nu}T\right)^{4}
-\frac{\nu}{1-2\nu}\,\left[r^{4(1-\nu)}-r^{2}\right]\,\Omega_K^0\,\rho_c^0\,,
\end{eqnarray}
where $r\equiv T/T_0$, $\,T_{0}\simeq 2.75\,K=2.37\times
10^{-4}\,eV\,$ is the present CMB temperature, $\,g_{\ast }=2\,$
for photons and $g_{\ast }=3.36\,$ if we take the neutrinos into
account, but this difference has no importance for the present
considerations.  It is easy to see that the size of the parameter
$\,\nu\,$ gets restricted, because for $\,\nu \geq 1\,$ the
density of radiation (in the flat case) would be the same or even
below the one at the present universe. On the other hand near the
nucleosynthesis time we have $T\gg T_0$, and so the corresponding
value of the CC is
\begin{eqnarray}\label{CCR2}
\Lambda_R(T)\simeq
\frac{\nu}{1-\nu}\,\frac{\pi^2\,g_{\ast}}{30}\,\left(r^{-\nu}T\right)^{4}-
\frac{\nu}{1-2\nu}\,\left[\frac{\nu}{1-\nu}\,r^{4(1-\nu)}-r^{2}\right]\,\Omega_K^0\,\rho_c^0\,.
\end{eqnarray}
Hence, in order not to be ruled out by the nucleosynthesis, the
ratio of the CC and the energy density at that time has to satisfy
\begin{equation}
|\,{\Lambda_R}\,/\,{\rho_R}\,| \,\simeq \,
|\,{\nu}\,/\,{(1-\nu)}\,| \,\simeq \, |{\nu}|\ll 1\,.
\label{ratioCCrho}
\end{equation}
A nontrivial range could e.g. be $\,\,0<|{\nu}|\leq 0.1$. Both
signs of $\nu$ are in principle allowed provided the absolute
value satisfies the previous constraint. Let us notice that, in
view of the definition (\ref{notations 1}), the condition $\nu\ll
1$ also means that $M\lesssim M_P$. Hence, the nucleosynthesis
constraint coincides with our general will to remain in the
framework of the effective approach. It is remarkable that the two
constraints which come from very different considerations, lead to
the very same restriction on the unique free parameter of the
model. The canonical choice $\,M^2=M_P^2$, corresponds to
\begin{equation}\label{nurange}
|\nu|=\nu_0\equiv\frac{1}{12\,\pi}\simeq  2.6\times 10^{-2}
\end{equation}
which certainly satisfies $\nu\ll 1$. Since the value of the
parameter $\,\nu\,$ is small, all the effects of the
renormalization group running of the CC will be approximately
linear in $\,\nu\,$.

\vspace{0.7cm}

 \noindent {\bf Some phenomenological Implications.}\quad

 \vspace{0.4cm}

Let us now turn to the ``recent'' Universe characterized by the
redshift interval $\,0 < z\lesssim 2$, and let us evaluate some
cosmological parameters which can be, in principle, extracted
from the future observations, say by the SNAP
project\,\cite{SNAP}. The first relevant exponent is the relative
deviation $\,\delta_{\Lambda}\,$ of the CC from the constant
value $\La_0$.  Then, using our solution (\ref{Lambda 1}), we
obtain (expanding at first order in $\nu$)
\begin{equation}\label{CCnusmall}
\CC(z)\simeq \CC_0+\nu\,\rho_M^0\,\left[(1+z)^3-1\right]\,.
\end{equation}

\begin{figure}[tb]
\vskip -0.5cm
\centerline{\resizebox*{10.5cm}{!}{\includegraphics{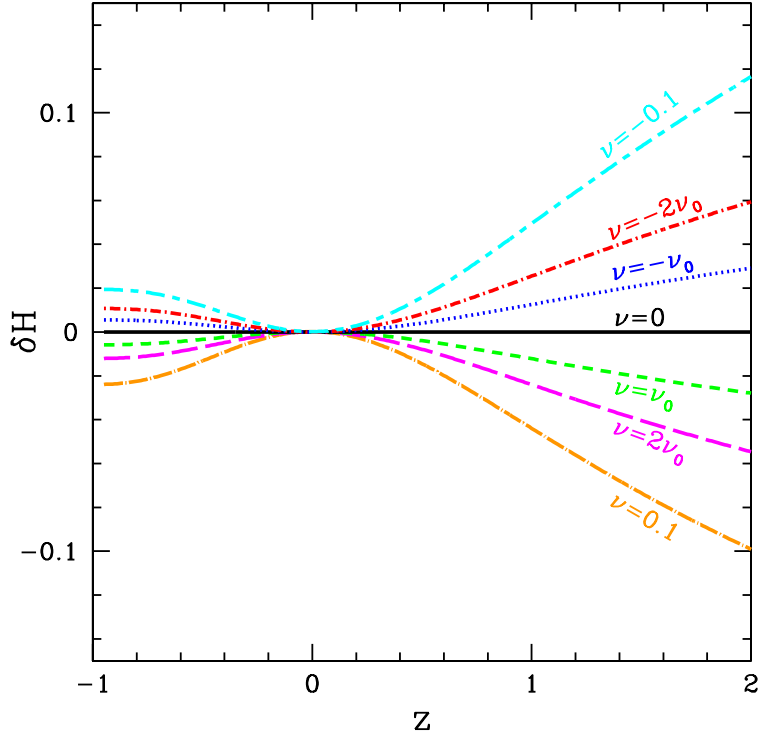}
} } \caption{Relative deviation $\delta H (z;\nu)$,
Eq.\,(\ref{deltaH}), of the Hubble parameter versus the redshift
for various values of the single free parameter $\nu$ in  flat
space with $\,\Omega_{M}^0=0.3$ and
$\,\Omega_{\Lambda}^0=0.7$.}\label{figure1}
\end{figure}

\begin{figure}[tb]
\vskip -0.5cm
\centerline{\resizebox*{10.5cm}{!}{\includegraphics{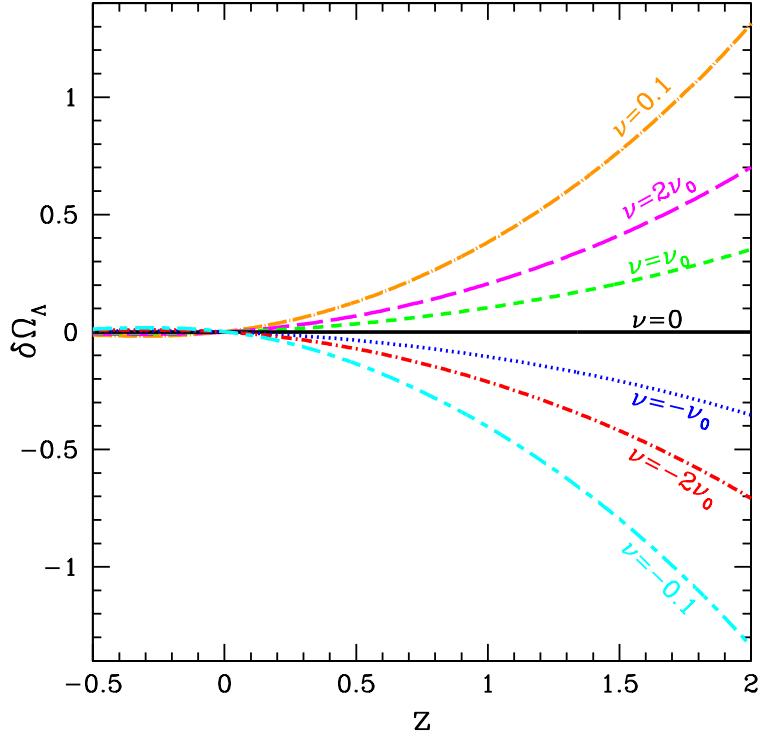}
} } \caption{Relative deviation $\delta\Omega_{\Lambda}(z;\nu)$,
Eq.(\ref{deviOmega}), of the cosmological constant parameter
versus the redshift for the same cases as in Fig.\,\ref{figure1}.}
\label{figure2}
\end{figure}
What about the numerical effect?  The relative correction to the
CC at redshift $z$ is given by
\begin{equation}\label{deviLambd}
\delta_{\CC}\equiv \frac{{\CC}(z;\nu)-{\CC}_0}{{\CC}_0}
=\nu\,\frac{\OM^0}{\OL^0}\,\left[(1+z)^3-1\right]\,.
\end{equation}
Let us take flat space with $\Omega_M^0=0.3$, $\OL^0=0.7$, and the
value $\nu=\nu_0$ defined in (\ref{nurange}). For $z=1.5$
(reachable by SNAP \cite{SNAP}) we find $\delta_{\CC}=16.3\%$,
and so one that should be perfectly measurable by SNAP. For
larger values of $\nu$ the test can be much more efficient of
course. For instance, values of order $\nu=0.1$ are still
perfectly tenable, in which case the previous correction would be
as large as $\gtrsim 60\%$! In general, the strong cubic
$z$-dependence in $\,\delta_{\Lambda}(z;\nu)\,$ should manifest
itself in the future CC observational experiments where the range
$\,z\gtrsim 1$ will be tested. It is important to emphasize that
$\,\nu\,$ is the unique arbitrary parameter of this model for a
variable CC. Therefore, the experimental verification of the
above formula must consist in: {\it i)} pinning down the sign and
value of the parameter $\,\nu$; and {\it ii)} fitting that
formula to the experimental data.

Next we present the relative deviation of the Hubble parameter
$\,H(z,\nu)\,$ with respect to the conventional one
$\,H(z,\nu=0)$. At first order in $\nu$,
\begin{eqnarray}\label{deltaH}
&&\delta
H(z;\nu)\equiv\,\frac{H(z;\nu)-H(z;0)}{H(z;0)}\nonumber\\
&\simeq &-\frac12\,\nu\,\Omega_M^0\,\frac{1+(1+z)^3\,
\left[3\ln(1+z)-1\right]}{1+\Omega_M^0\,\left[(1+z)^3-1\right]}\,.
\end{eqnarray}
Equation (\ref{deltaH}) gives the leading quantum correction to
the Hubble parameter (\ref{Friedmann}) when the renormalization
effects in (\ref{rho}) and (\ref{Lambda 1}) are taken into
account. Notice that $\delta H(0;\nu)=0$, because for all $\nu$
we have the same initial conditions. Then for $z\neq 0$ we have
e.g.\, $\delta H (1.5;\nu_0)\simeq\,-2\%$ and $\delta
H(2;\nu_0)\,\simeq\,-3\%$. For higher $\nu$ we get sizeable
effects: $\delta H(z;0.1)\simeq\,-8\%$ and $-10\%\,$ for $z=1.5$
and $z=2$ respectively. See Fig.\,\ref{figure1} for the detailed
numerical evaluation.

Although the induced corrections on $H$ are not very high, the
relative deviation of the renormalized cosmological constant
parameter
$\Omega_{\Lambda}(z;\nu)=8\pi\,G\Lambda(z;\nu)/3H^2(z;\nu)$ with
respect to the standard one turns out to be much more significant.
We get at leading order in $\nu$,
\begin{eqnarray}\label{deviOmega}
&& \delta {\Omega_{\Lambda}}(z;\nu)\equiv
\frac{\Omega_{\Lambda}(z;\nu)
-\Omega_{\Lambda}(z;0)}{\Omega_{\Lambda}(z;0)}\\
&&\simeq\nu\,\left[\frac{{\Omega}_{M}^0\,(1+z)^3-1}{\Omega_{\Lambda}^0}
 +\frac{1+3\,{\Omega}_{M}^0\,(1+z)^3\,\ln(1+z)}
{\Omega_{\Lambda}^0+{\Omega}_{M}^0\,(1+z)^3}\right]\nonumber\,.
\end{eqnarray}

Again $\,\,\delta\Omega_{\Lambda}(0;\nu)=0$, as it should.
Moreover, the deviation has the two expected limits for the
infinite past and future, viz.
$\delta\Omega_{\Lambda}(\infty;\nu)=\infty$ and
$\delta\Omega_{\Lambda}(-1;\nu)=0$. The numerical evaluation of
(\ref{deviOmega}) is given in Fig. \ref{figure2}.  Notice that
even for $\nu$ as small as $\nu_0$, (\ref{nurange}), there is a
sizeable $20\%$ increase of $\Omega_{\Lambda}$  at redshift
$z=1.5$ -- reachable by SNAP. For $\nu=2\,\nu_0$ the increase at
$z=1.5$ is huge, $\sim 40\%$.\ For $\nu<0$ the effects go in the
opposite direction. For fixed $\nu$, the larger is $z$ the larger
are the quantum effects. Thus if $\nu\simeq\nu_0$ and some
experiment in the future can reach the far $z=2$ region with
enough statistics, then the effects on $\Omega_{\Lambda}(z)$ and
the Hubble parameter can be dramatic. If they are seen, this
model can provide an explanation for them.  At present
$\Omega_{\Lambda}$ has been determined at roughly $10\%$ from
both supernovae and CMB measurements, and in the future SNAP will
clinch $\Omega_{\Lambda}$ to within $\pm 0.05$\,\cite{SNAP}. The
previous numbers show that for $z\gtrsim 1$ the cosmological
quantum corrections can be measured already for a modest
$\nu\gtrsim 10^{-2}$.  A complete numerical analysis of this kind
of FLRW models, including both the flat and curved space cases,
together with a detailed comparison with the present and future
Type Ia supernovae data, will be presented
elsewhere\,\cite{BigOne}.


\vspace{0.5cm}

 \noindent {\bf Conclusions}\quad

 \vspace{0.3cm}

To summarize, we have presented a semiclassical FLRW type of
cosmological model based on a running CC with the scale $\mu=H$.
We have shown that if the decoupling quantum effects on $\Lambda$
have the usual form as for the massive fields, then we can get a
handle on problem II by giving an alternative answer to problem
III. In fact, we can explain the variation of the CC at low
energies without resorting to any scalar field. The CC is mainly
driven, without fine tuning, by the ``relic'' quantum effects from
the physics of the highest available scale (the Planck scale), and
its value naturally lies in the acceptable range. It is remarkable
that all relevant information about the unknown world of the high
energy physics is accumulated into a single parameter $\,\nu$.
Finally, we have shown that the next generation of supernovae
experiments, like SNAP, should be sensitive to $\nu$ within its
allowed range. A non-vanishing value of $\nu$ produces a cubic
dependence of the CC on $z$ at high redshift, which should be well
measurable by that experiment, if it is really there. Therefore an
efficient check of this alternative framework, which might be the
effective behavior common to a large class of high energy
theories, can in principle be performed in the near future.

\vspace{0.7cm}

 \noindent {\bf Acknowledgments.}\quad

 \vspace{0.3cm}

I.Sh. is thankful to E.V. Gorbar for useful discussions.
The work of I.Sh. has been supported by the research grant from
FAPEMIG (Minas Gerais, Brazil) and by the fellowship from CNPq
(Brazil). The work of J.S. has been supported in part by MECYT
and FEDER under project FPA2001-3598, and also by the Dep. de
Recerca de la Generalitat de Catalunya under contract
2002BEAI400036. J.S. also thanks the warmth hospitality at the
Dep. de Fisica, U. F. Juiz de Fora. C.E.B and P.R.L are supported
by grants AYA2000--0983 and RTN2-2001--0037.


\end{document}